\documentclass{article}
\usepackage{parskip}
\usepackage{float}
\usepackage{amsmath}
\usepackage{multirow}
\usepackage[table]{xcolor}
\usepackage{array}

\usepackage[english]{babel}

\usepackage[a4paper,top=2cm,bottom=2cm,left=3cm,right=3cm,marginparwidth=1.75cm]{geometry}

\usepackage{amsmath}
\usepackage{graphicx}
\usepackage[colorlinks=true, allcolors=blue]{hyperref}

\title{How to design a network architecture using capacity planning}
\author{Gilbert Moïsio \\ Network \& Methodology Senior Consultant \\ gmoisio@gmail.com}
\date{February 20, 2016}

\begin{document}
\maketitle

\begin{abstract}
Building a network architecture must answer to organization needs, but also to two major elements which are the need for dependability and performance. By performance, we must understand the ability to meet an immediate need and the ability to scale without reducing the performance of the whole as new elements are added to the network infrastructure. This last point is covered by Capacity Planning domain.
\end{abstract}

\section{Introduction}

For a very long time, network architectures have been inspired by the model called Three-Tier Architecture. This model is composed of three levels: Core, Distribution and Access. Server farms are just a special case of the Access level. The Distribution and Access levels form Building Blocks which are concentrated on the Core level. The objective is to size Building Blocks that will not grow beyond a threshold established in advance. Increasing the size of the network requires the addition of Building Blocks. The sizing of the Building Blocks and the connections between all the layers of the architecture uses the notion called over-subscription.
As explained in \cite{1}, statistical considerations play a key role in the field of network Capacity Planning. It is intuitively obvious that, in the worst case, the results of provisioning lead to unnecessary waste and expense. It is also intuitively obvious that in a network with over-subscription there is a risk of not being able to provide adequate services to users if the ratio that is used is too aggressive.
Network Capacity Planning is a field often based on empirical observations, trends, patterns and forecasts of network usage. In a domain subject to bursty traffic, it is difficult to derive precise link sizing parameters and it is common to see networks designed for worst-case usage scenarios. However, it is precisely the bursty nature of traffic that allows a statistical approach to be used in Capacity Planning. In other words, we exploit the fact that not everyone is online at the same time to make our network sizing decisions.

\section{Mathematical approach}

As proposed in \cite{1}, it is possible to approach the subject from a statistical point of view. Assume a number \emph{n} of \emph{on/off} type traffic sources that can use the network links. Each source generates a burst of traffic of up to \emph{R} (bits per second) in a time interval of \emph{T} seconds during its \emph{on} period, then that source goes \emph{off} for another period of \emph{T} seconds, and finally the cycle repeats itself. Note that the source sends \emph{RT} bits during the \emph{on} period and that the number of bits per period is random.
The different sources are not synchronized in time (they can go \emph{on} and \emph{off} independently of each other) and their aggregated traffic is processed by the network links.
We can compute the statistical over-subscription for all the sources, using the following formula: \[C=\frac{R}{2}n+C_{\epsilon}S_{max}\sqrt{n}\]

\pagebreak

Into which:

\begin{itemize}
    \item \emph{C} is the capacity of the link to forecast in bits/s, if half of the sources are \emph{on} at the same time.
    \item $nR$ corresponds to the maximum traffic in bits/s, if all the sources are \emph{on} at the same time.
    \item $\frac{R}{2}n$ is the average traffic in bits/s, from all sources aggregated assuming that only half are \emph{on} at any one time.
    \item $C_{\epsilon}$ corresponds to the QoS level which reflects the confidence interval with which it can be declared that the resulting capacity will not be exceeded by the traffic load.
    \item $\epsilon=0.01$ corresponds to a confidence interval of $99\%$ or $1-\epsilon$, which gives $C_{\epsilon}=2.575829303549$ according to the Normal Law table.
    \item $S_{max}=\frac{R}{2\sqrt{3}}$ is the standard deviation in bits/s for a traffic source, where standard deviation is $\sqrt{variance}$.
\end{itemize}

Consider an example where $n=100$, $R=1,000,000\,bit/s$, $T=1\,sec$ and $\epsilon=0.01$.

\begin{align}
	C_{max}&=nR  \nonumber\\
	&=100\times1.000.000\,bits/s  \nonumber\\
	&=100\,Mbits/s  \nonumber
\end{align}

\begin{align}
    S_{max}&=\frac{R}{2\sqrt{3}}  \nonumber\\
    &=\frac{1.000.000\,bits/s}{3,46}  \nonumber\\
    &=289.017\,bits/s  \nonumber
\end{align}

\begin{align}
    C_{stat}&=\frac{1.000.000\,bits/s}{2}\times100+2,58\times289.017\,bits/s\times10 \nonumber\\
    &=50\,Mbits/s+7,5\,Mbits/s \nonumber\\
    &=57,5\,Mbits/s  \nonumber
\end{align}

In this example, we can see that in the worst case provisioning, the link would require \emph{100 Mbps} of capacity, while a link with capacity of around \emph{60 Mbps} would be sufficient to support the volume of traffic, with a confidence level of $99\%$ (i.e. we are convinced that $99\%$ of the time, the total volume of traffic will not exceed \emph{60 Mbits/s}).

\section{Ratios approach}

\begin{figure}[H]
\centering
\includegraphics[width=0.8\textwidth]{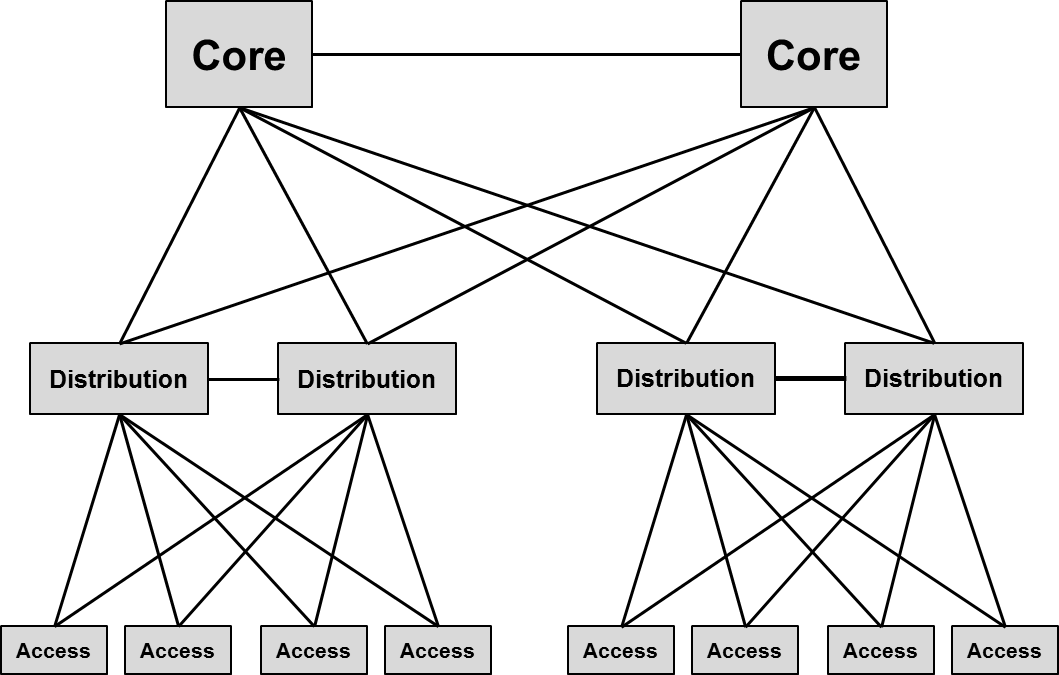}
\caption{\label{fig:ThreeTier}Three-Tier architecture}
\end{figure}

In a faster and more pragmatic way, good practices propose over-subscription ratios of \emph{20:1} for the links between the Access level and the Distribution level, of \emph{4:1} for the links between the Distribution level and the Core level and \emph{1:1} for the links between the Access level of the server farms and the Core level. If congestion should occur on the interconnection links, the frames are queued and it is the QoS that can take over in order to prioritize critical flows.
If these rules are not respected, the algorithms used by the TCP protocol may see segment losses which will not allow them to take full advantage of the speeds available.

If it was always relatively obvious that the interconnection links between the Access level and Distribution level were at level 2, the links between the Distribution level and the Core level could be at level 2 or at level 3. The point of setting up routing was to ensure that all the links were used in parallel by seeking the shortest path or by distributing the load over all the paths of equivalent cost. In this case, setting up Layer 2 virtual networks through the architecture was less easy to achieve, although not impossible. The use of level 2 for the links between the Distribution level and the Core level made the architecture simpler, but did not make it possible to use all the links in active mode, since a protocol for avoiding Ethernet loop, such as the Spanning Tree would deactivate the links corresponding to the redundant paths.

The arrival of the aggregation of the Control Plane between several switches (stacking, virtualization, ...) has changed the vision of the three-level architecture, into which the Spanning Tree protocol no longer sees a redundant path. The duplicated links going up on two switches, whose control plane is merged, are aggregated together using the LACP IEEE 802.3ad protocol.

\begin{figure}[H]
\centering
\includegraphics[width=0.8\textwidth]{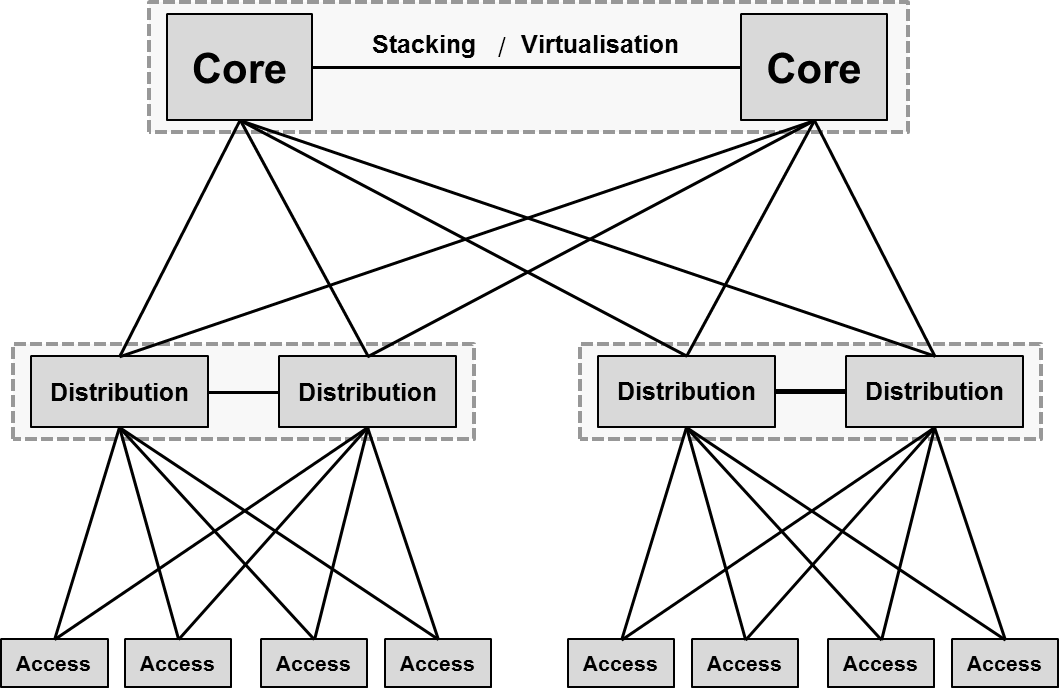}
\caption{\label{fig:ThreeTierAggregation}Three-Tier architecture with aggregated Control Plane}
\end{figure}

Not only are all the links active, which increases speeds and promotes over-subscription ratios, but in the event of the loss of one of them, the re-convergence times of the architecture can reach less than a second. This Subsecond Convergence architecture is more suited to the transport of UDP streams with strong constraints such as voice.

The consequence of changing application deployment, increased use of virtual machines, and redesigned storage has resulted in changing traffic patterns in the particular building block of server farms from predominantly client/ server (north-south) to a significant level of server-to-server (east-west) flow. These changes in the flow matrix led to the rebirth, under the name of Leaf-Spine, of the architecture developed by Charles Clos within Bell Laboratories in the 1950s. In his document \cite{2}, Charles Clos introduces the concept of a multi-stage switched network, the advantage of which is to allow connection between a large number of input and output ports with small intermediate switches. The mathematical model makes it possible to realize a totally non-blocking network like a crossbar switch. The demonstration is usually done with a three-stage system (Ingress Stage or Input Switches, Middle Stage or Intermediary Switches, and Egress Stage or Output Switches), but the Leaf-Spine model we use is represented on Two-Tier (as Charles Clos pointed out when entry points are also exit points), into which the Spine tier is the aggregation level and the Leaf tier represents the Access level.

\begin{figure}[H]
\centering
\includegraphics[width=0.8\textwidth]{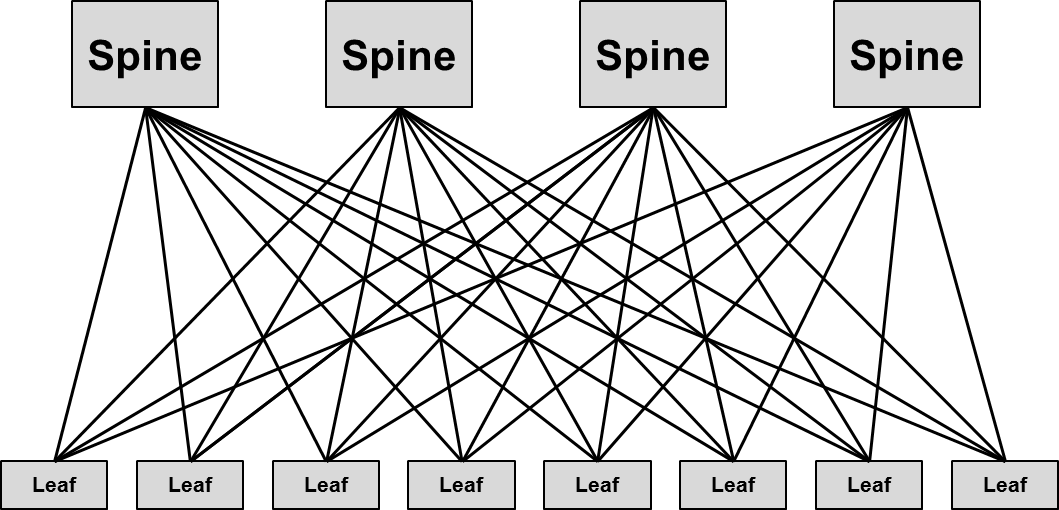}
\caption{\label{fig:LeafSpine}Leaf-Spine architecture}
\end{figure}

In this architecture, it is no longer necessary to merge the Control Plane of the Spine switches, but all the links must be active for the model to work. In order not to be subject to the use of a level 2 loop avoidance protocol such as the Spanning Tree, a dynamic routing protocol such as OSPF (Open Shortest Path First) makes it possible to create an architecture whose links are used in ECMP (Equal-Cost MultiPath). Routing protocols such as IS-IS (Intermediate System to Intermediate System) or BGP (Border Gateway Protocol) can also be implemented in this kind of architecture. We find the difficulty of propagation of level 2 virtual network through the whole architecture, which is circumvented by the encapsulation of Ethernet frames in UDP thanks to the use of the VXLAN protocol (Virtual eXtensible LAN – RFC 7348) which achieves a MAC-in-IP tunnel for the transport of level 2 virtual networks on a level 3 network (concept called Overlay Network as opposed to the supporting level 3 network which is called Underlay Network). The NVGRE protocol (Network Virtualization using Generic Routing Encapsulation – RFC 7637) is another Overlay technology allowing to build a layer 2 network on an infrastructure configured in layer 3.

The SPB (Shortest Path Bridging – IEEE 802.1aq) and TRILL (TRansparent Interconnection of Lots of Links – RFC 6325 corrected by RFCs 6327, 6439, 7172, 7177, 7357, 7179, 7180, 7455, 7780 and 7783) protocols have been proposed as an alternative to the Spanning Tree to be able to achieve this architecture entirely in level 2 with all the active links. Manufacturers make it possible to build this type of level 2 architecture for small to medium-sized networks.

\begin{figure}[H]
\centering
\includegraphics[width=0.8\textwidth]{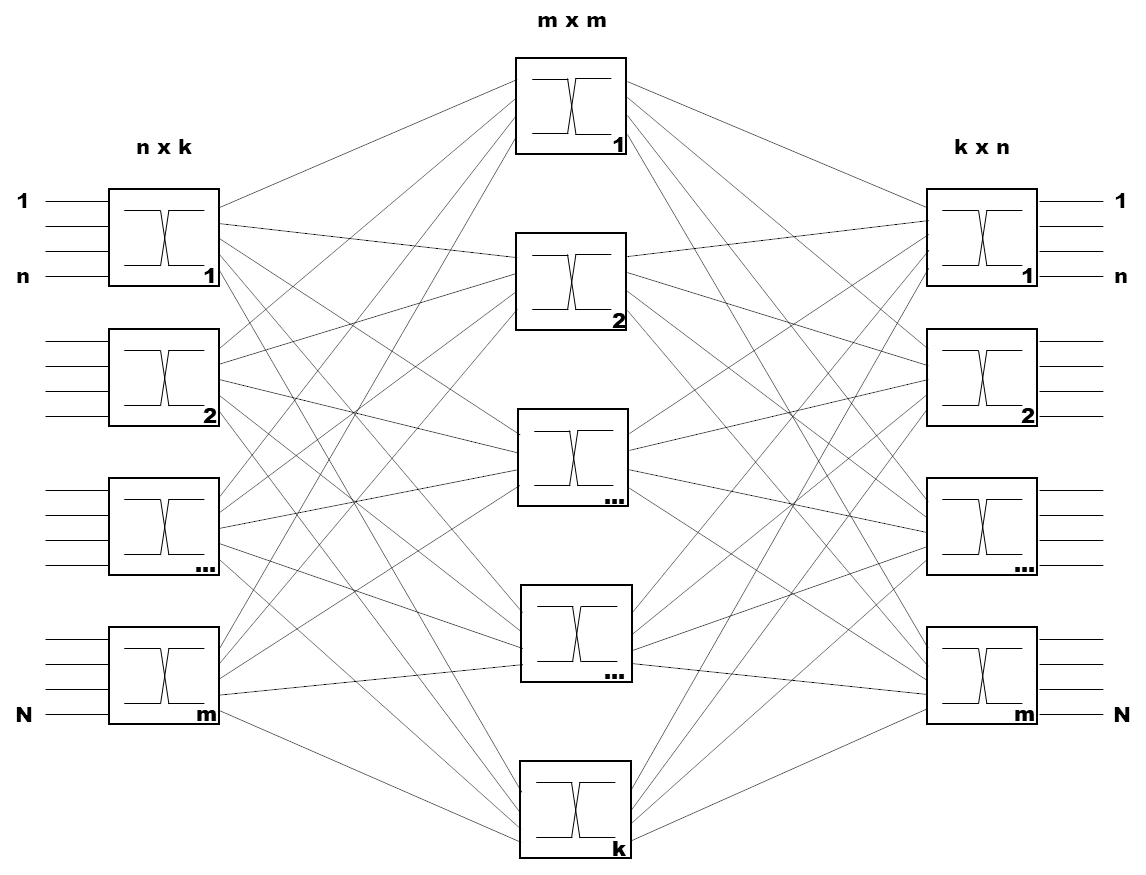}
\caption{\label{fig:ClosNetwork}3-stage Clos Network}
\end{figure}

The mathematical model can be extract from the Standford University course \cite{3}. It can be shown that with $k \geq n$, the Clos network can be non-blocking like a crossbar switch.

For the Leaf-Spine model to be non-blocking, it would be necessary that for each pair of ports, there is a path arrangement to connect the input port with the output port, which would require the use of a large amount of spine nodes and interconnecting links. At the Spine level, the over-subscription ratio is generally \emph{1:1}, i.e. the speed of the aggregate links of a Leaf switch is equal to that of another Leaf switch so that that all received traffic can be resent in a non-blocking fashion. The over-subscription ratio of a leaf switch to a spine switch must be planned according to needs and it is generally considered that a ratio of \emph{3:1} is acceptable. This could be write: \[k \geq \frac{n}{3}\]
As all the input and output links of a leaf can have different speeds, it is necessary to include variables to represent them, using \emph{i} as output speed and \emph{j} as input speed: \[ik \geq \frac{jn}{3}\]

\section{Ethernet performance background}

The IEEE (Institute of Electrical and Electronics Engineers) 802.3 Working Group develops standards for Ethernet networks. A number of active projects focus on different throughputs. A standard Ethernet frame consists of the following fields:

\begin{center}
\begin{tabular}{ l c c }
    \textbf{Frame Part} &	\textbf{Minimum Frame Size} &	\textbf{Maximum Frame Size} \\
    Inter Frame Gap (96 ns)\footnotemark[1] &	12 Bytes &	12 Bytes \\
    MAC Preamble (+SFD) &	8 Bytes &	8 Bytes \\
    MAC Destination Address &	6 Bytes &	6 Bytes \\
    MAC Source Address &	6 Bytes &	6 Bytes \\
    MAC Type (or length) &	2 Bytes &	2 Bytes \\
    Payload (Network PDU) &	46 Bytes &	1500 Bytes \\
    Check Sequence (CRC) &	4 Bytes &	4 Bytes \\
    \textbf{Total Frame Physical Size} &	\textbf{84 Bytes} &	\textbf{1,538 Bytes}
\end{tabular}
\end{center}

\footnotetext[1]{96 ns for Gigabit Ethernet and 9.6 ns for 10 Gigabit Ethernet}

The maximum capacity of an interface in frames per second, or speed, is calculated using the following formula: \[Maximum Number of Frames per Second = \frac{Ethernet Data Rate (bits per second)}{Total Frame Physical Size (bits)}\]

If we take Gigabit Ethernet as an example, we get the following performance:

\begin{center}
\begin{tabular}{ c c }
    \begin{tabular}{@{}c@{}}\textbf{Maximum Rate with} \\ \textbf{Minimum Frame Size}\end{tabular} &	\begin{tabular}{@{}c@{}}\textbf{Maximum Rate with} \\ \textbf{Maximum Frame Size}\end{tabular} \\
    $\frac{1,000,000,000 \, b/s}{(84 \, B \times 8 \, b/B)} = 1,488,095 \, f/s$ &	$\frac{1,000,000,000 \, b/s}{(1,538 \, B \times 8 \, b/B)} = 81,274 \, f/s$
\end{tabular}
\end{center}

To compute the throughput of an Ethernet interface, the Inter Frame Gap and MAC Preamble fields are not taken into account, as they do not constitute useful information. The Check Sequence field may or may not be included in the calculation depending on how you see it. Therefore, our example of Gigabit Ethernet becomes:

\begin{center}
\begin{tabular}{ l c c }
    \begin{tabular}{@{}l@{}}\textbf{Maximum Gigabit} \\ \textbf{Bandwidth}\end{tabular} &	\textbf{With CRC (4 Bytes)} &	\textbf{Without CRC} \\
    \begin{tabular}{@{}l@{}}Minimum Frame Size \\ (60 Bytes)\end{tabular} & \begin{tabular}{@{}c@{}}$1,488,095 \, f/s \times (64 \, B \times 8 \, b/B)$ \\ $\approx 762 \, Mbps$\end{tabular}  & \begin{tabular}{@{}c@{}}$1,488,095 \, f/s \times (60 \, B \times 8 \, b/B)$ \\  $\approx 714 \, Mbps$\end{tabular} \\
    \begin{tabular}{@{}l@{}}Maximum Frame Size \\ (1514 Bytes)\end{tabular} & \begin{tabular}{c}$81,274 \, f/s \times (1,518 \, B \times 8 \, b/B)$ \\ $\approx 987 \, Mbps$\end{tabular}  & \begin{tabular}{c}$81,274 \, f/s \times (1,514 \, B \times 8 \, b/B)$ \\ $\approx 984 \, Mbps$\end{tabular} 
\end{tabular}
\end{center}

Applied to 10 Gigabit Ethernet technology we can generates the following figures:

\begin{center}
\begin{tabular}{ c c }
    \begin{tabular}{@{}c@{}}\textbf{Maximum Rate with} \\ \textbf{Minimum Frame Size}\end{tabular} &	\begin{tabular}{@{}c@{}}\textbf{Maximum Rate with} \\ \textbf{Maximum Frame Size}\end{tabular} \\
    $14,880,952 \, f/s$ &	$812,743 \, f/s$
\end{tabular}
\end{center}

\begin{center}
\begin{tabular}{ l c c }
    \textbf{Maximum 10 Gigabit Bandwidth} &	\textbf{With CRC (4 Bytes)} &	\textbf{Without CRC} \\
    Minimum Frame Size (60 Bytes) & $\approx 7.62 \, Gbps$  & $\approx 7.14 \, Gbps$ \\
    Maximum Frame Size (1514 Bytes) & $\approx 9.87 \, Gbps$  & $\approx  9.84 \, Gbps$ 
\end{tabular}
\end{center}

The size of an Ethernet frame can be increased by at least one Tag Id of 4 bytes due to the use of 802.1Q vlan. It is also possible to use a larger frame size, called Jumbo Frame, which can reach a total size of 9038 bytes.

\pagebreak

\section{TCP and UDP performance background}

The Payload field of the Ethernet frame contains the data that has been encapsulated in TCP or UDP, then in IP.

\begin{center}
\begin{tabular}{ l c c c c }
    \textbf{Frame Component} &&	\textbf{TCP} &&	\textbf{UDP} \\
    Inter Frame Gap &&	12 Bytes &&	12 Bytes \\
    MAC Preamble (+SFD) &&	8 Bytes &&	8 Bytes \\
    MAC Destination Address &&	6 Bytes &&	6 Bytes \\
    MAC Source Address &&	6 Bytes &&	6 Bytes \\
    MAC Type (or length) &&	2 Bytes &&	2 Bytes \\
    \rowcolor{lightgray}
    Payload (Network PDU) & IPv4 Header	&	20 Bytes & IPv4 Header &	20 Bytes \\
    \rowcolor{lightgray}
    46 - 1500 Bytes & TCP Header	&	20 Bytes & UDP Header &	8 Bytes \\
    \rowcolor{lightgray}
    & \begin{tabular}{@{}c@{}}TCP options \& \\ Data/Padding\end{tabular}	&	\begin{tabular}{@{}c@{}}6 – 1460 \\ Bytes\end{tabular} & Data/Padding &	\begin{tabular}{@{}c@{}}18 – 1472 \\ Bytes\end{tabular} \\
    Check Sequence (CRC) &&	4 Bytes &&	4 Bytes \\
    \textbf{Total Frame Physical Size} & \multicolumn{4}{c}{\textbf{84 - 1,538 Bytes}}
\end{tabular}
\end{center}

\begin{center}
\begin{tabular}{ l c }
    \textbf{Gigabit Ethernet TCP/IP \& UDP/IP Throughput} & \\
    \begin{tabular}{@{}l@{}}Max TCP/IP Data Rate (84 Bytes Frames) and \\ Min TCP/IP Packet (60 Bytes + 4 Bytes CRC)\end{tabular}	& \begin{tabular}{@{}c@{}}$1,488,095 \, f/s \times 6 \, B/f \times 8 \, b/B$ \\ $\approx 71 \, Mbps$\end{tabular} \\
    \begin{tabular}{@{}l@{}}Max TCP/IP Data Rate (1538 Bytes Frames) and \\ Max TCP/IP Packet (1514 Bytes + 4 Bytes CRC)\end{tabular}	& \begin{tabular}{@{}c@{}}$81,274 \, f/s \times 1,448 \, B/f \times 8 \, b/B$ \\ $\approx 941 \, Mbps$\end{tabular} \\
    \textbf{With TCP/IP TimeStamp} & \\
    \begin{tabular}{@{}l@{}}Max TCP/IP Data Rate (1538 Bytes Frames) and \\ Max TCP/IP Packet (1514 Bytes + 4 Bytes CRC)\end{tabular}	& \begin{tabular}{@{}c@{}}$81,274 \, f/s \times 1,460 \, B/f \times \, 8 b/B$ \\ $\approx 949 \, Mbps$\end{tabular} \\
    \textbf{Without TCP/IP TimeStamp} & \\
    \begin{tabular}{@{}l@{}}Max UDP/IP Data Rate (84 Bytes Frames) and \\ Min UDP/IP Packet (60 Bytes + 4 Bytes CRC)\end{tabular}	& \begin{tabular}{@{}c@{}}$1,488,095 \, f/s \times 18 \, B/f \times 8 \, b/B$ \\ $\approx 214 \, Mbps$\end{tabular} \\
    \begin{tabular}{@{}l@{}}Max UDP/IP Data Rate (1538 Bytes Frames) and \\ Max UDP/IP Packet (1514 Bytes + 4 Bytes CRC)\end{tabular}	& \begin{tabular}{@{}c@{}}$81,274 \, f/s \times 1,472 \, B/f \times \, 8 b/B$ \\ $\approx 957 \, Mbps$\end{tabular} \\
\end{tabular}
\end{center}

TCP (Transmission Control Protocol) and UDP (User Datagram Protocol) are used to transfer data or packets over networks. TCP is connection oriented while UDP is connectionless. TCP transmission is more reliable, but slower because it checks for errors and maintains data order. UDP does not have an error checking mechanism that is why it is less reliable but faster in data transfer.

There are several implementations of TCP (Tahoe, Reno, Vegas, New Reno…), but fundamentally the operation of TCP remains the same and is based on RFC 5681. It is worth knowing the implementation used on the system and change it if it turns out that it is not the most suitable for the intended use.

\begin{center}
\begin{tabular} { l c c l }
    \textbf{TCP variant} &	\textbf{Network} &	\textbf{Class} & \textbf{Main features} \\
    Reno & Standard & Loss-based & Standard TCP \\
    Vegas & Standard & Delay-based & \begin{tabular}{@{}l@{}}Proactive scheme \\ RTT and rate estimation\end{tabular} \\
    Veno & Wireless & Delay-based & \begin{tabular}{@{}l@{}}Combine Reno and Vegas \\ Deal with random loss\end{tabular} \\
    Westwood & Wireless & Delay-based & Deal with « Large » dynamic channels \\
    BIC & Long fat\footnotemark[2] & Loss-based & Modification of congestion avoidance scheme \\
    CUBIC & Long fat & Loss-based & Improved variant of BIC \\
    HSTCP & Long fat & Loss-based & Modification of congestion avoidance scheme \\
    Hybla & Long fat & Delay-based & Modification of congestion avoidance scheme \\
    Scalable & Long fat & Loss-based & Modification of congestion avoidance scheme \\
    Illinois & Long fat & Loss-Delay-based & Modification of congestion avoidance scheme \\
    YeAH & Long fat & Delay-based & \begin{tabular}{@{}l@{}}Two working modes for the congestion \\ avoidance phase: fast and slow\end{tabular} \\
    HTCP & Long fat & Delay-based & Modification of congestion avoidance scheme \\
    LP & - & Delay-based & Variant for low priority flows \\
\end{tabular}
\end{center}    

\footnotetext[2]{High Latency}

\pagebreak

There are four fundamental congestion control algorithms: slow start, congestion avoidance, fast retransmit and fast recovery. The slow start and congestion avoidance algorithms must be used by the sender to control the amount of data injected into the network. To implement these algorithms, two variables are used: \emph{cwnd} (sender side congestion window) and \emph{rwnd} (receiver's advertised window). The \emph{cwnd} variable is the amount of data the sender can transmit over the network before receiving an acknowledgment (ACK) and the \emph{rwnd} variable is the limit on the receiver side. It is the minimum of these two variables which is taken as a reference in the data transmission. Another state variable: \emph{ssthresh} (slow start threshold) is used to determine which of the two algorithms slow start or congestion avoidance should be used to control data transmission.

The slow start algorithm is used at startup or after a loss detected by the retransmission timer, to determine the available network capacity in order to avoid congestion. \emph{IW} (Initial Window) is the initial value of \emph{cwnd} and should be set using the following rules:

\begin{itemize}
    \item If $Sender \, MSS > 2190 \, bytes$ then $IW = 2 * SMSS \, bytes$ and must not be greater than 2 segments
    \item If $(SMSS > 1095 \, bytes) \, and \, (SMSS \leq 2190 \, bytes)$ then $IW = 3 * SMSS \, bytes$ and must not be more than 3 segments
    \item If $SMSS \leq 1095 \, bytes$ then $IW = 4 * SMSS \, bytes$ and should not be more than 4 segments
\end{itemize}

The initial value of \emph{ssthresh} should be set to the size of the largest possible window (advertised window), but \emph{ssthresh} should be reduced in response to congestion. The slow start algorithm is used when $cwnd < ssthresh$, while the congestion avoidance algorithm is used when $cwnd > ssthresh$. In the event of a tie, the issuer can choose which algorithm to use. During the slow start phase, TCP increments \emph{cwnd} by at most \emph{SMSS bytes} and slow start stops when \emph{cwnd} reaches or exceeds \emph{ssthresh}. It is recommended to augment \emph{cwnd} in the following way: \[cwnd \mathrel{+}= min(N, SMSS)\] with \emph{N} being the number of bytes that were acknowledged in the last ACK.

During the congestion avoidance phase, \emph{cwnd} is increased by approximately one segment per RTT (Round-Trip Time) and the algorithm continues its work until congestion is detected. It is recommended to increase \emph{cwnd} as follows and adjust each time an ACK is received: \[cwnd \mathrel{+}= SMSS \times \frac{SMSS}{cwnd}\]

When a transmitter detects the loss of a segment using the retransmission timer and this segment has not been retransmitted, the value of \emph{ssthresh} must not be set beyond the value given by the equation: \[ssthresh=max(\frac{FlightSize}{2}, 2\times SMSS)\] with \emph{FlightSize} being the amount of data waiting in the network.

A receiver, in TCP, should immediately send a duplicate ACK when an out-of-order segment arrives, with the objective of informing the sender of what happened and requesting sending the segment again. The sender should use the fast retransmit algorithm to detect and repair the loss based on incoming duplicate ACKs. The algorithm uses the arrival of three duplicate ACKs as an indication that a segment has been lost. From then on, TCP performs a retransmission of the lost segment without waiting for the expiration of the retransmission timer. After transmission of the lost segment, the fast recovery algorithm takes over the transmission of new data until a non-duplicate ACK arrives. The implementation of the fast retransmit and fast recovery algorithms follows rules for positioning values for \emph{cwnd} and \emph{ssthresh}.

In a simplified way, the TCP throughput is calculated with the following formula: \[Throughput = \frac{Window Size}{RTT}\]

Studies have shown that it is necessary to take into account other parameters such as MSS (Maximum Segment Size) and packet loss. As explained in \cite{4}, assuming that the network has a packet loss probability called \emph{p}, the sender will be able to send an average of $\frac{1}{p}$ packets before a packet loss. According to this model, \emph{cwnd} will never exceed a maximum \emph{W} (window expressing the number of segments that can be sent during an RTT), because at approximately $\frac{1}{p}$ packets, a new packet loss will cause the division by two from \emph{cwnd}. Therefore, the total quantity of data delivered at each cycle of $\frac{1}{p}$ packets is given by: \[\frac{1}{p} = \left( \frac{W}{2} \right) ^2 + \frac{1}{2} \left( \frac{W}{2} \right) ^2 = \frac{3}{8} W^2\]

Therefore, $MSS \times \frac{3}{8} W^2$ bytes are emitted every $RTT \times \frac{W}{2}$ cycle.

As $W = \sqrt{\frac{8}{3p}}$ the bit rate is therefore:

\begin{align}
	Throughput &= \frac{MSS \times \frac{3}{8} W^2}{RTT \times \frac{W}{2}}\nonumber\\
	&= \frac{\frac{MSS}{p}}{RTT \sqrt {\frac{2}{3p}}}\nonumber\\
	&= \frac{MSS \times \sqrt{\frac{3}{2}}}{RTT \times \sqrt{p}}\nonumber\\
	&= \frac{MSS \times 1.22}{RTT \times \sqrt{p }}\nonumber
\end{align}

A study \cite{5} compared different TCP implementations and highlighted the performance differences in situations where RTT and loss probability values vary. As a result, on a wired network, the TCP Hybla and TCP CUBIC implementations achieve very good performance while TCP Reno also performs well, but TCP-LP achieves the worst results. On a wireless network, the cards are redistributed and the TCP Reno, TCP BIC and TCP-LP implementations obtain the best results. On long distance networks with delay, such as satellite networks, the TCP CUBIC implementation performs best. Finally, the implementation that seems to behave homogeneously in all circumstances remains TCP Reno.

\bibliographystyle{alpha}
\bibliography{sample}

\end{document}